\documentclass[conference]{IEEEtran}
\pdfoutput=1
\IEEEoverridecommandlockouts
\usepackage{cite}
\usepackage{amsmath,amssymb,amsfonts}
\usepackage{graphicx}
\usepackage{textcomp}
\usepackage{xcolor}
\usepackage{subfig}

\begin{document}

\title{Deep data compression for approximate ultrasonic image formation\\
\thanks{This work was supported by Applus+ RTD, CWI and the Dutch Research Council (NWO 613.009.106, 639.073.506). Submission - IEEE International Ultrasonics Symposium 2020.}}
\author{
    \IEEEauthorblockN{Georgios Pilikos\IEEEauthorrefmark{1}, Lars Horchens\IEEEauthorrefmark{2}, Kees Joost Batenburg\IEEEauthorrefmark{1}\IEEEauthorrefmark{3}, Tristan van Leeuwen\IEEEauthorrefmark{1}\IEEEauthorrefmark{10} and Felix Lucka\IEEEauthorrefmark{1}\IEEEauthorrefmark{4}}
    \IEEEauthorblockA{\IEEEauthorrefmark{1}Computational Imaging, Centrum Wiskunde \& Informatica, Amsterdam, NL
    }
    \IEEEauthorblockA{\IEEEauthorrefmark{2}Applus+ E\&I Technology Centre, Rotterdam, NL}
    \IEEEauthorblockA{\IEEEauthorrefmark{3}Leiden Institute of Advanced Computer Science, Leiden University, Leiden, NL}
    \IEEEauthorblockA{\IEEEauthorrefmark{10}Mathematical Institute, Utrecht University, Utrecht, NL}
    \IEEEauthorblockA{\IEEEauthorrefmark{4}Centre for Medical Image Computing, University College London, London, UK}
}

\maketitle

\begin{abstract}
In many ultrasonic imaging systems, data acquisition and image formation are performed on separate computing devices. Data transmission is becoming a bottleneck, thus, efficient data compression is essential. Compression rates can be improved by considering the fact that many image formation methods rely on approximations of wave-matter interactions, and only use the corresponding part of the data. Tailored data compression could exploit this, but extracting the useful part of the data efficiently is not always trivial. In this work, we tackle this problem using deep neural networks, optimized to preserve the image quality of a particular image formation method. The Delay-And-Sum (DAS) algorithm is examined which is used in reflectivity-based ultrasonic imaging. We propose a novel encoder-decoder architecture with vector quantization and formulate image formation as a network layer for end-to-end training. Experiments demonstrate that our proposed data compression tailored for a specific image formation method obtains significantly better results as opposed to compression agnostic to subsequent imaging. We maintain high image quality at much higher compression rates than the theoretical lossless compression rate derived from the rank of the linear imaging operator. This demonstrates the great potential of deep ultrasonic data compression tailored for a specific image formation method.
\end{abstract}


\section{Introduction}
Ultrasonic imaging is becoming faster, more portable and is being utilized in remote locations\cite{ultrafast}. This leads to many clinical and industrial applications where data acquisition and ultrasonic image formation are performed on separate computing devices. Often, data transfer is performed across very limited capacity channels. Combined with the rise of 3D ultrasonic imaging and the increasing amount of data, fast ultrasonic data transfer is becoming a technical barrier.

There have been efforts to improve data transfer by acquiring less data without compromising image quality using compressed sensing (CS) \cite{csus}, \cite{cs2} and finite rate of innovation (FRI) approaches \cite{xampling}. However, CS relies on randomly weighted combinations of acquired data which might not always be possible in hardware. Slow sparse reconstruction algorithms could also compromise the real-time requirement. In addition, assumptions about the data are needed (e.g. sparsity) for both CS and FRI approaches that do not always hold in practice.

Recently, deep learning methods have been introduced for ultrasonic imaging using a limited amount of data \cite{huijben} - \cite{segmNair}. In addition, deep learning has been applied for ultrasonic data compression and decompression \cite{perdios_compression}. Data reduction rates obtained by generic compression schemes can be further improved using a key idea from our work. That is, we explicitly take the end goal into account, which is to form an image of sufficient quality to guide decision making processes. This is beneficial since many ultrasonic image formation methods rely on approximations of the acoustic wave-matter interactions. This translates to using only the corresponding part of the data which is a small fraction of the original. However, extracting the useful part of the data efficiently is not always trivial.

In this work, we tackle this problem using deep neural networks. We propose a novel encoder-decoder architecture with vector quantization in the bottleneck and explicitly include the image formation as a network layer. It results in a data-to-image compression optimized to preserve the image quality of a particular image formation method. Here, we examine the Delay-And-Sum (DAS) algorithm used in reflectivity-based imaging. In section 2, we describe the ultrasonic data acquistion and image formation. Then, in section 3, we introduce our proposed data-to-image compression. In section 4, we include experiments on simulated data and compare our approach with data-to-data compression, agnostic to image formation. 

\begin{figure*}
\centering
\includegraphics[scale=0.1153]{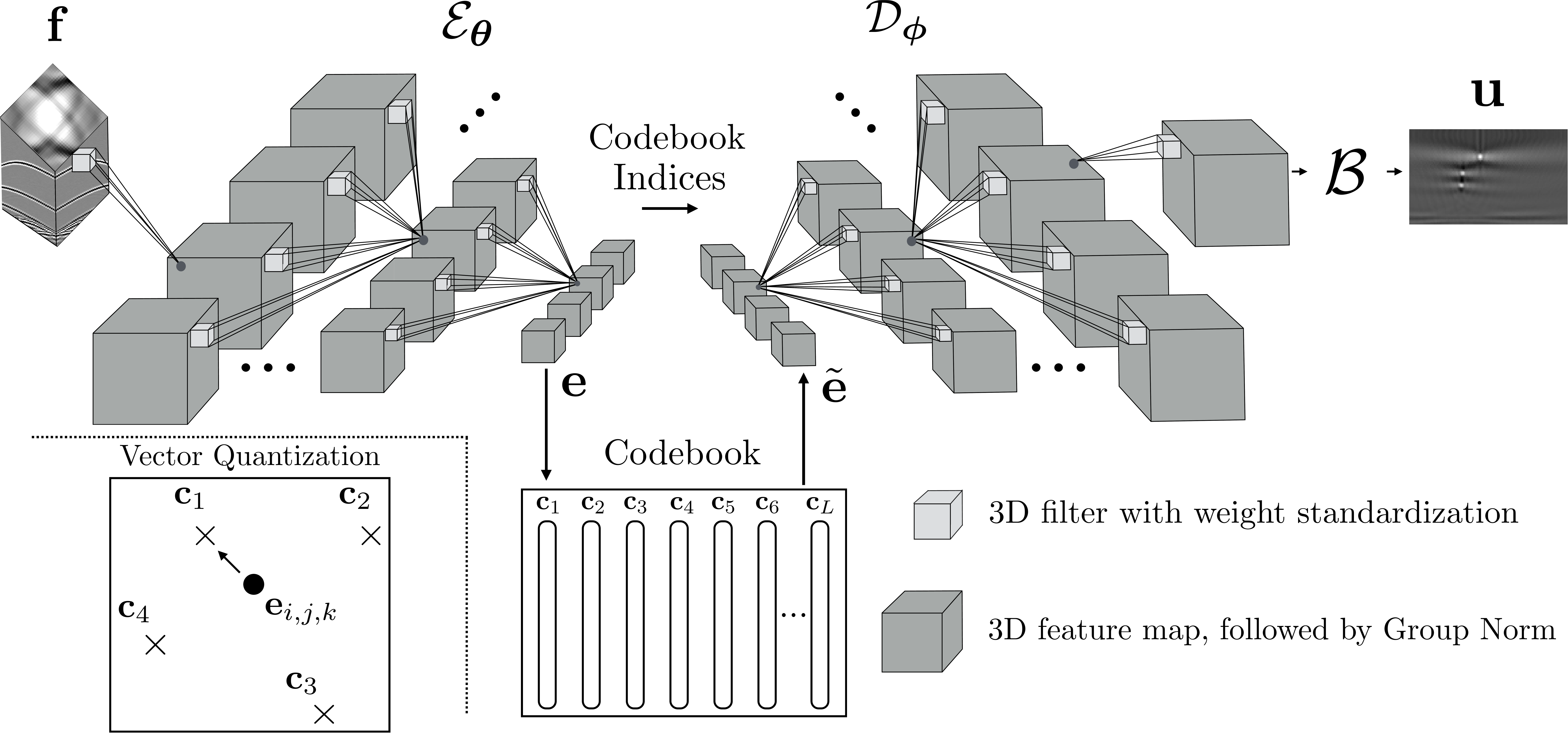}
\caption{The encoder is a 3D DCNN, $\mathcal{E}_{\boldsymbol \theta}$, that uses striding to downsample feature maps. At the bottleneck, a vector quantization layer is used to compress the code further. A shared codebook maps the output of the encoder, $\mathbf e$, to the closest codes via a nearest neighbour search. A 2D depiction is included for illustration purposes in the bottom left corner. The indices to these codes are transmitted and the feature maps, $\tilde{\mathbf e}$, are recovered. These are passed to the decoder, $\mathcal{D}_{\boldsymbol \phi}$, which has a similar, albeit in reverse, architecture to that of the encoder. It is followed by the DAS operator, $\mathcal B$. Feature maps are followed by Group Norm and ReLU (tanh is used at the last feature map of decoder). Only one filter at one location per layer is illustrated.}
\label{d2i}
\end{figure*}

\section{Ultrasonic imaging}
During data acquisition, a pulsed ultrasonic wave is transmitted into a medium of interest using source at location $\mathbf r_s$. Receivers at locations $\mathbf r_m$ capture the resulting wave field which contains information about the medium's acoustic properties. Data acquisition continues with the next source firing until all elements have been used as sources, which leads to data, $\mathbf f \in \mathbb{R}^{n_t \times n_s \times n_r}$ \cite{tfm}, \cite{iwex}. The number of time samples, sources and receivers is $n_t$, $n_s$ and $n_r$ respectively. The aim is to obtain an image, $\mathbf u \in \mathbb{R}^{n_x \times n_z}$, with $n_x$ and $n_z$ being the pixels in horizontal and vertical directions. 

We consider the DAS image formation which calculates travel times, $\tau(\mathbf p_i, \mathbf r_{s}, \mathbf r_{m})$, between each source $\mathbf r_{s}$, image point, $\mathbf p_i$ and receiver,  $\mathbf r_{m}$.
It performs a \emph{delay} operation in the data and a \emph{sum} across all combinations of travel times. Thus,
\begin{equation}
u_i = \sum_{s=0}^{n_s} \sum_ {m=0}^{n_r} f(\tau(\mathbf p_i, \mathbf r_{s}, \mathbf r_{m}),s,m),
\end{equation} which is repeated for all image points to form an image. The whole process can be written as,
\begin{equation}\label{das}
\mathbf u = \mathcal{B}\mathbf{f},
\end{equation} where $\mathcal{B}: \mathbb{R}^{n_t \times n_s \times n_r}  \to  \mathbb{R}^{n_x 
\times n_z} $ is a linear operator. We will refer to this as \emph{DAS operator} hereafter. As DAS corresponds to a linear operator, the best lossless linear compression is given by the projection onto its row-space. The corresponding compression rate is the ratio between the size of the data and the rank of $\mathcal{B}$. However, this projection is difficult to compute and apply in practice.

\section{Proposed data-to-image compression}
To achieve practical and high data compression rates, we propose a novel encoder-decoder architecture with a vector quantization layer in the bottleneck and explicitly incorporate the DAS operator as an image forming network layer. Figure \ref{d2i} depicts the proposed data-to-image compression architecture.

The encoder is composed of a 3D deep convolutional neural network (DCNN) with varying number of layers depending on the desired compression rate. Each layer has 32 filters except for the last layer which has 256. Each filter has $5\times 5\times 5$ dimensions. We use a stride of 2 at various layers to downsample the feature maps for compression. The encoder defines a mapping, $\mathcal E_{\boldsymbol \theta}$, and its output is given by,
\begin{equation}
\mathbf e = \mathcal E_{\boldsymbol \theta}(\mathbf f),
\end{equation} where $\mathbf e \in \mathbb{R}^{n_1 \times n_2 \times n_3 \times D}$. $n_1$, $n_2$ and $n_3$ are the dimensions of the reduced data and $D=256$ is the number of filters in the last layer of the encoder. The encoder's compression rate depends on the dimensions of these feature maps which depend on the stride, filter dimensions and number of layers. We can compress this latent space further using vector quantization.

Inspired by the Vector Quantised Variational AutoEncoder (VQ-VAE) \cite{vqvae}, we use a vector quantization layer at the output of the encoder. A shared codebook, $\mathbf C \in \mathbb{R}^{D \times L}$, is introduced which is composed of entries, $\{ \mathbf c_1, \mathbf c_2, ..., \mathbf c_L\}$, with each code, $\mathbf c_l \in \mathbb{R}^{D}$ and the corresponding index set, $\mathcal I = \{1,...,L\}$. The number of codes in the codebook we use, L, is $512$. The output of the vector quantization layer is,
\begin{equation}
\mathbf q = Q^{\downarrow}_{\mathbf C}(\mathbf e),
\end{equation} with $\mathbf q \in \mathcal{I}^{n_1 \times n_2 \times n_3}$ and
\begin{equation}
q_{i,j,k} = \underset{v \in \mathcal{I}}{\operatorname{argmin}} \| \mathbf e_{i,j,k} - \mathbf c_v \|_{2},
\end{equation} where $\mathbf e_{i,j,k} \in \mathbb{R}^D$. This is a nearest neighbour search resulting in a 3D latent code, $\mathbf q$, with each entry corresponding to an index. This can be viewed as a non-linearity that maps the output of the encoder to 1-of-L vectors. A 2D depiction can be seen at the bottom left corner of Figure \ref{d2i}. 

The feature maps, $\tilde{\mathbf e} \in \mathbb{R}^{n_1 \times n_2 \times n_3 \times D}$, are recovered using the indices, $\mathbf q$, and the shared codebook as,
\begin{equation}
\tilde{\mathbf e} = Q^{\uparrow}_{\mathbf C}(\mathbf q),
\end{equation} where each entry corresponds to a code, $\tilde{\mathbf e}_{i,j,k} = \mathbf c_{q_{i,j,k}}$. The decoder, $\mathcal{D}_{\boldsymbol \phi}$, uses these recovered feature maps and outputs, 
\begin{equation}
\tilde{\mathbf f} = \mathcal{D}_{\boldsymbol \phi}(\tilde{\mathbf e}),
\end{equation} where $\tilde{\mathbf f} \in \mathbb{R}^{n_t \times n_s \times n_r}$. Its architecture resembles that of the encoder in reverse (using upsampling). Subsequently, we include the image formation, $\mathcal{B}$, into the network by implementing a layer that applies the DAS algorithm to input data and the adjoint of this operation to images to enable end-to-end training using back propagation. Skip connections are used in the encoder and decoder in intermediate layers to enable better information flow and reduce training time \cite{unet}. Weight Standardization \cite{ws} and Group Normalization \cite{gn} is used which helps training stability.

\subsection*{Loss function and training strategies}
Training involves the learning of parameters, $\{ \boldsymbol \theta, \mathbf C, \boldsymbol \phi\}$, for the encoder, the shared codebook, and the decoder. To achieve this, we would like to optimize
\begin{equation}
(\boldsymbol{\skew{2}\hat\theta}, \hat{\mathbf C}, \boldsymbol{\skew{3}\hat\phi}) = \underset{(\boldsymbol \theta, \mathbf C, \boldsymbol \phi)}{\operatorname{argmin}} \sum_{i} \| \mathbf u^{(i)} - \mathcal{B} \mathcal{D}_{\boldsymbol \phi}( Q^{\uparrow}_{\mathbf C}(Q^{\downarrow}_{\mathbf C}(\mathcal{E}_{\boldsymbol \theta}(\mathbf f^{(i)}) ) ) ) \|_{2}^2.
\end{equation} Unfortunately, $Q^{\uparrow}_{\mathbf C}$ and $Q^{\downarrow}_{\mathbf C}$ are not differentiable which means that at the encoder/decoder bottleneck, there is no real gradient. We approximate it using the straight-through estimator \cite{vqvae}. This translates to just copying the gradients from the decoder input to the encoder output. These gradients contain information as to how the encoder should change its output. However, the straight-through estimator does not allow any gradients to flow through the codebook. To learn the codebook space, we add a term in the loss to move the codebook entries, $\mathbf c_l$, closer to $\mathcal{E}_{\boldsymbol \theta}(\mathbf f)$. Given that the codebook entries could change without any constraint, we would like to enforce the encoder to commit to a codebook. Thus, a commitment loss term is also necessary. The overall loss function is given by,
\begin{multline}\label{loss}
(\boldsymbol{\skew{2}\hat\theta}, \hat{\mathbf C}, \boldsymbol{\skew{3}\hat\phi}) = \underset{(\boldsymbol \theta, \mathbf C, \boldsymbol \phi)}{\operatorname{argmin}} \sum_{i} \| \mathbf u^{(i)} - \mathcal{B} \mathcal{D}_{\boldsymbol \phi}(\mathcal{E}_{\boldsymbol \theta}(\mathbf f^{(i)}) ) \|_{2}^2 + \\ \sum_{i} \|\text{sg}(\mathcal{E}_{\boldsymbol \theta}(\mathbf f^{(i)})) - \mathbf C \|^{2}_{2} + \sum_{i} \| \mathcal{E}_{\boldsymbol \theta}(\mathbf f^{(i)}) - \text{sg}(\mathbf C) \|^{2}_{2},
\end{multline} where sg() is the stop-gradient operator. It is defined as identity in the forward pass and it has zero derivatives. The first term of the loss function is the data misfit which optimizes the weights for both the encoder and the decoder. The second term optimizes the codebook space only and the third term optimizes the weights of the encoder only.

\begin{figure*}
\centering
\subfloat[]{ 
	\includegraphics[scale=0.3647]{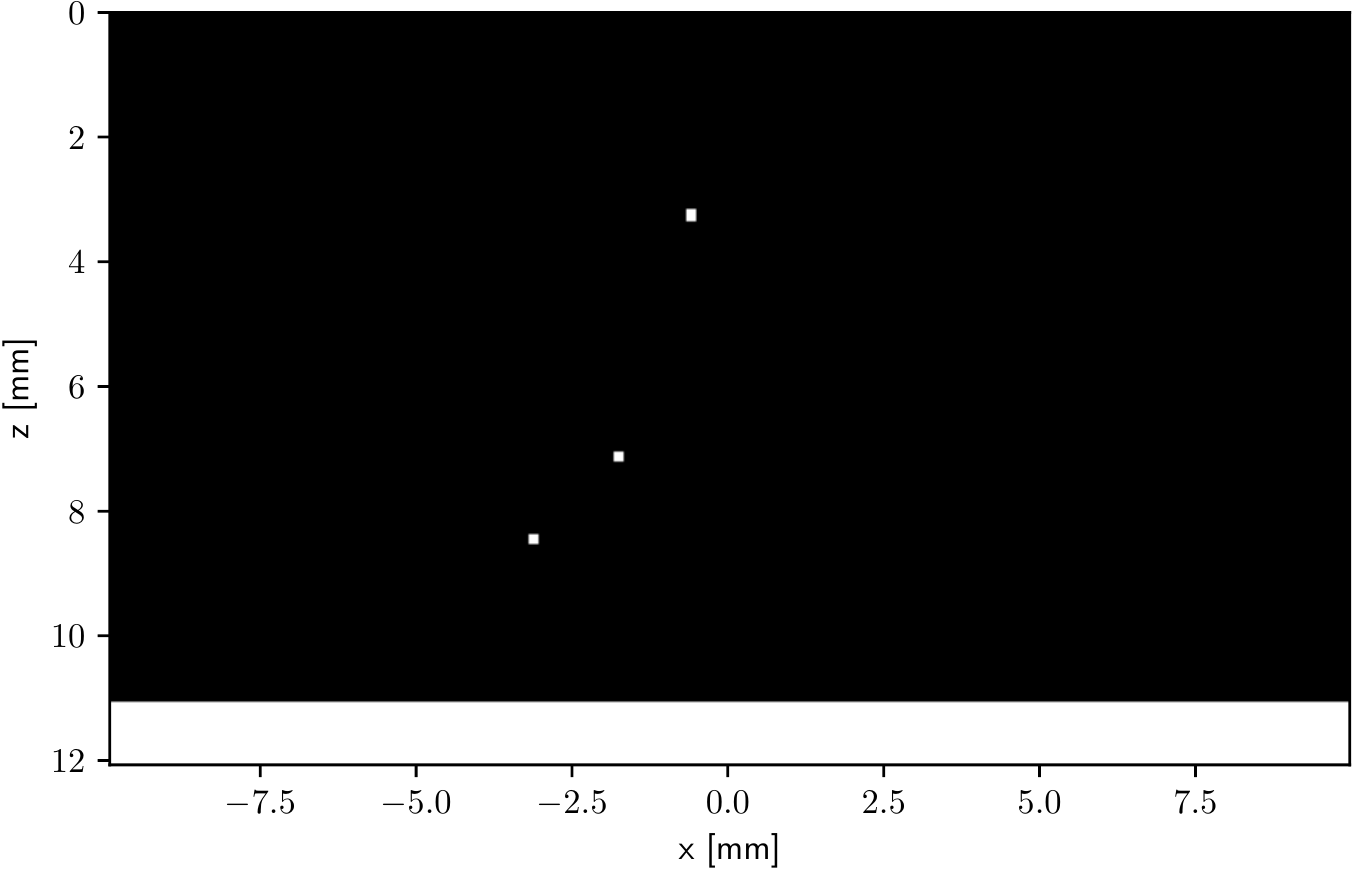}
	}
\subfloat[]{ 
	\includegraphics[scale=0.3647]{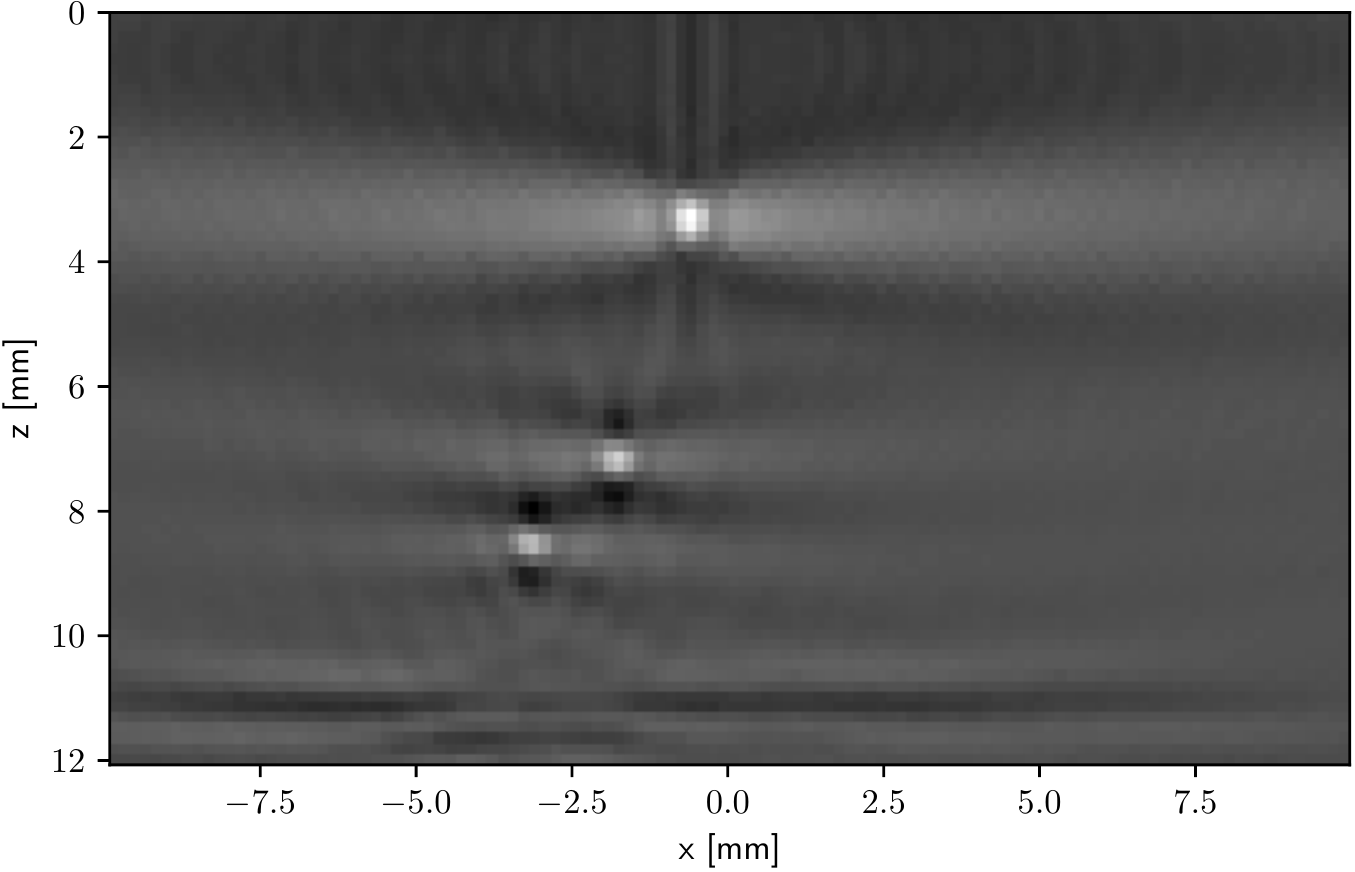}
	}
\subfloat[]{ 
	\includegraphics[scale=0.3647]{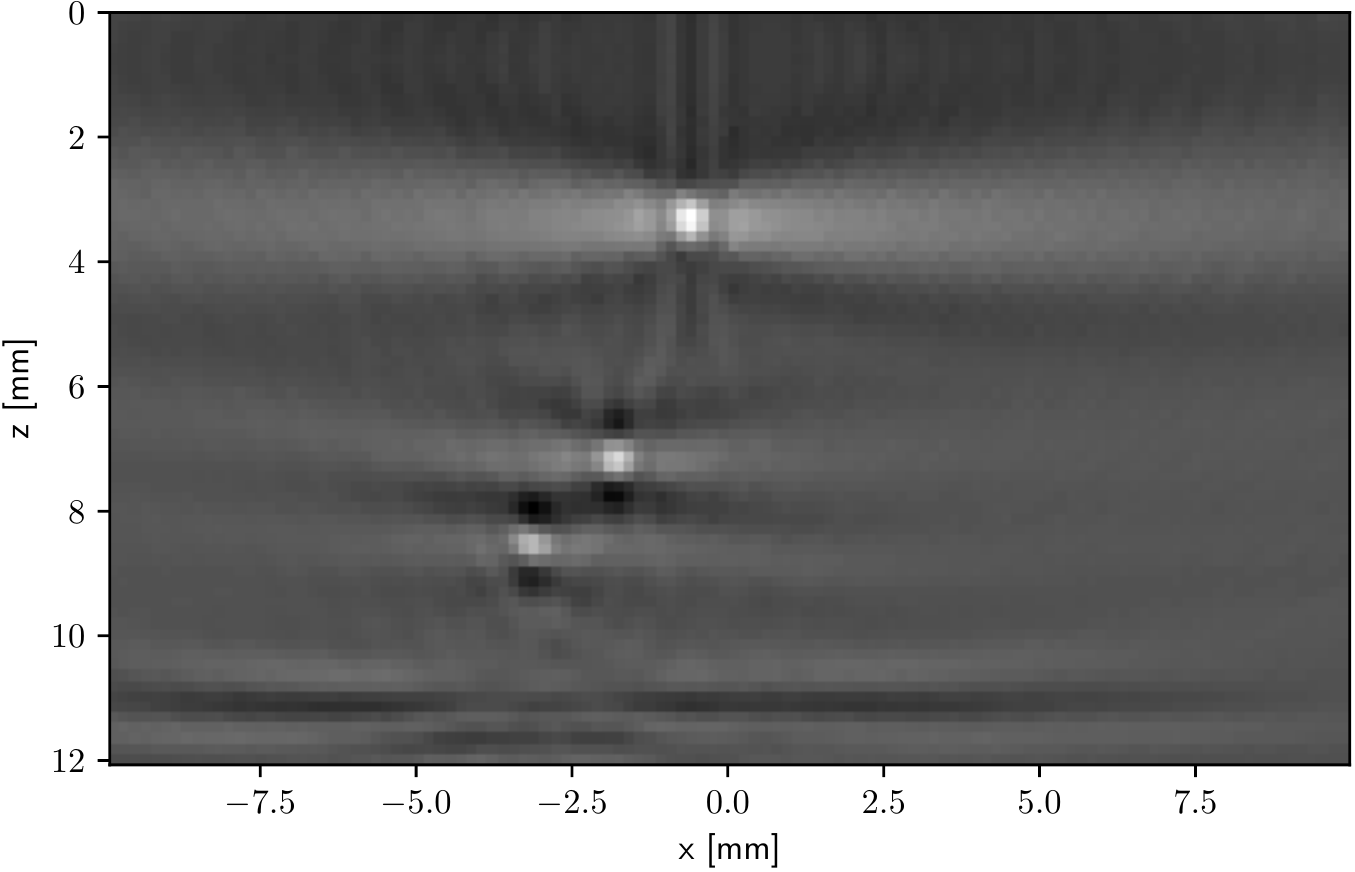}
	} \\
\subfloat[]{ 
	\includegraphics[scale=0.3647]{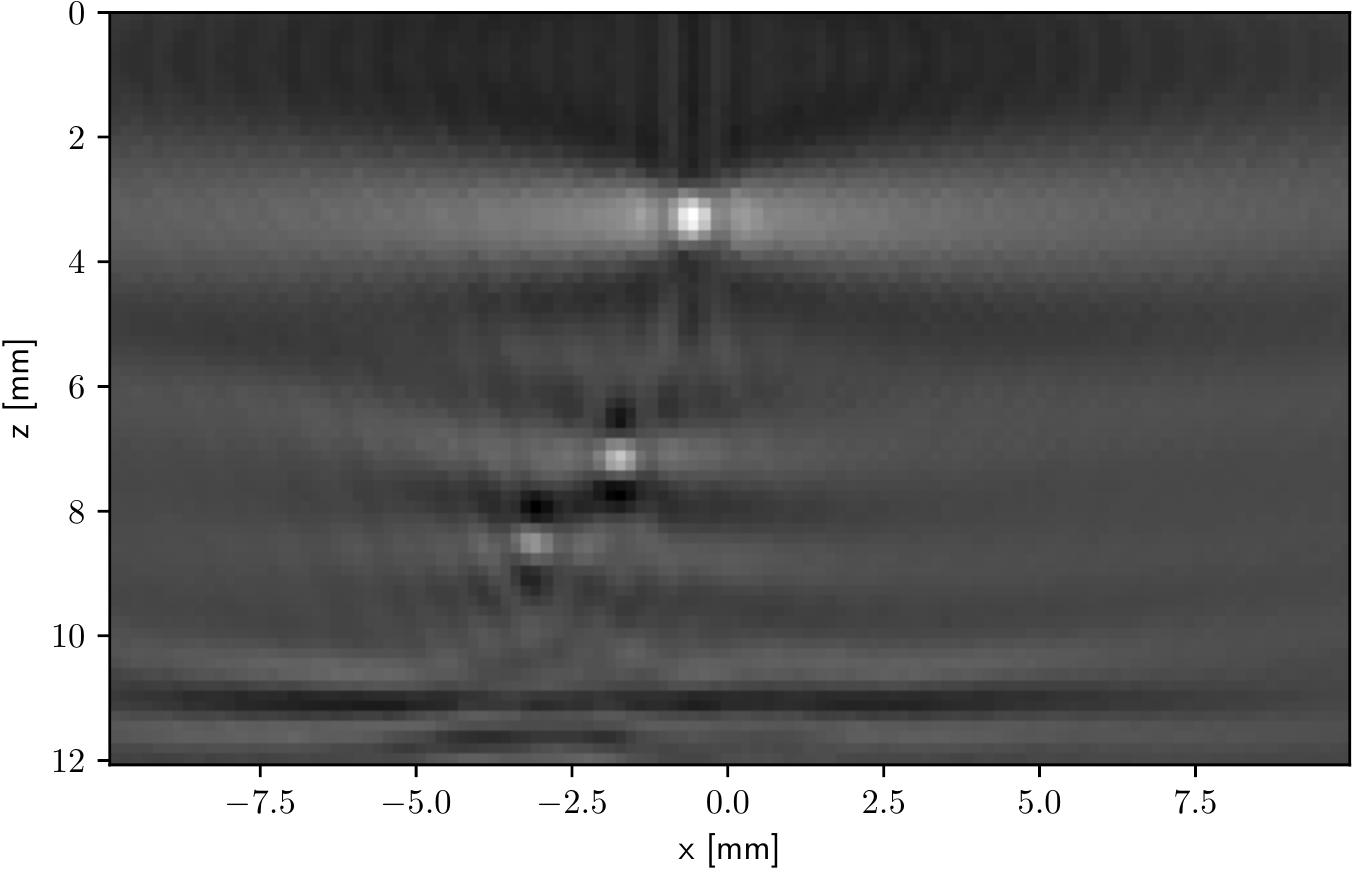}
	}	 
\subfloat[]{ 
	\includegraphics[scale=0.3647]{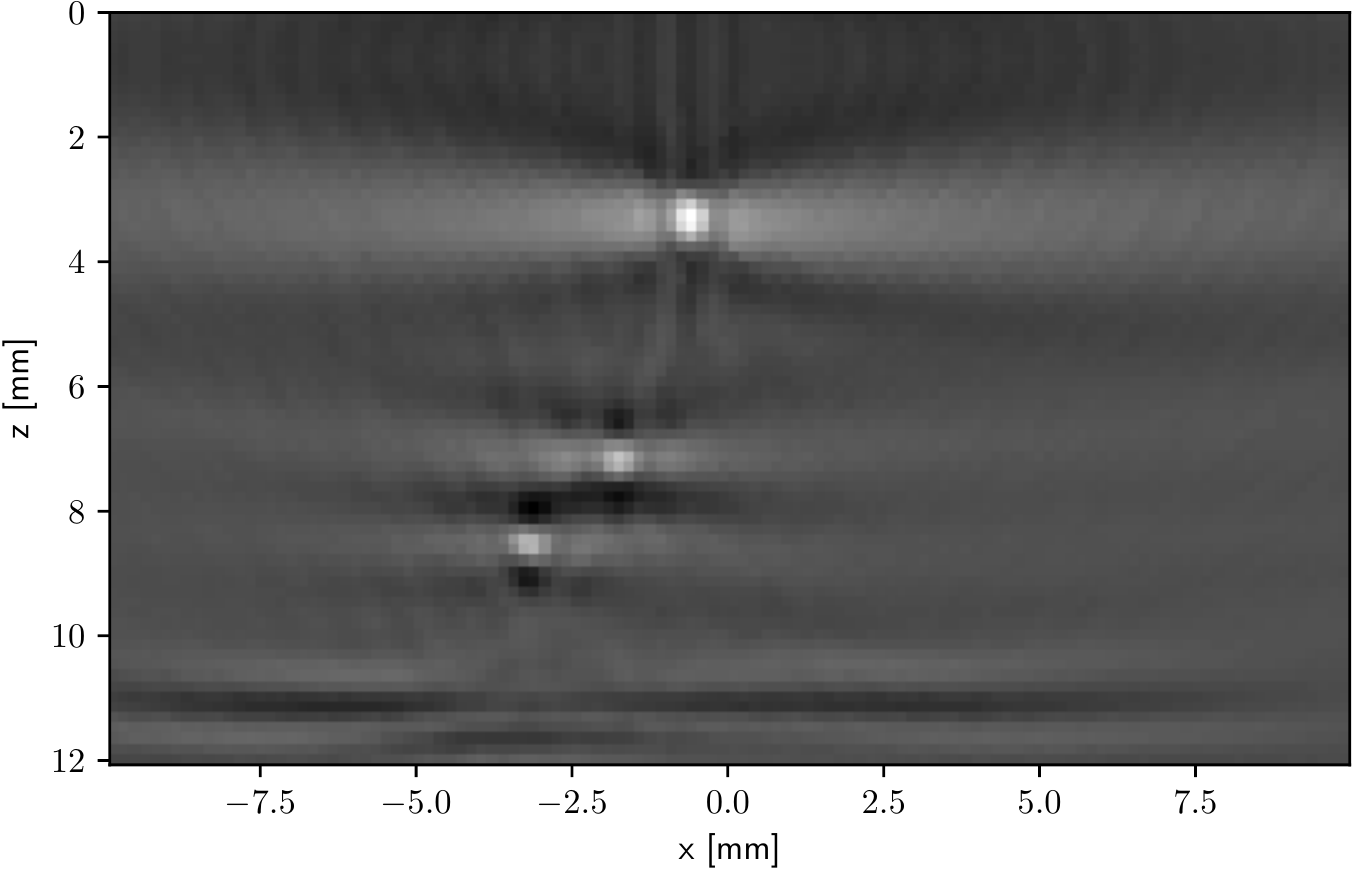}
	}
\subfloat[]{ 
	\includegraphics[scale=0.3647]{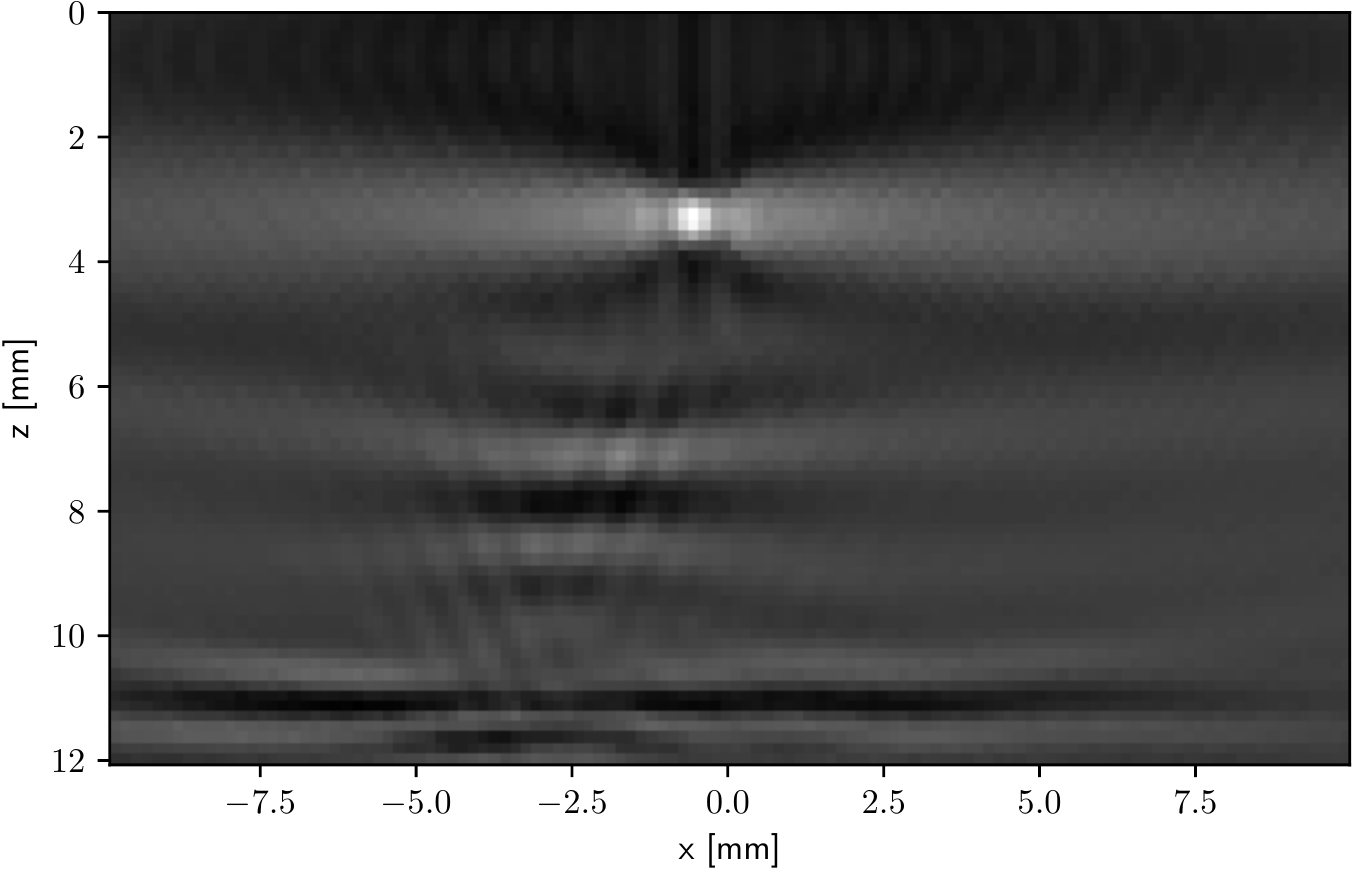}
	} 	
\caption{Two different compression rates are shown using two compression strategies. (a) Speed of sound map (white: air, black: carbon steel), (b) DAS image from original data, (c) data-to-image compression (SSIM = $0.94$, compression rate $\approx 468$), (d) data-to-data compression followed by DAS (SSIM = $0.85$, compression rate $\approx468$), (e) data-to-image compression (SSIM = $0.92$, compression rate $\approx1393$), (f) data-to-data compression followed by DAS (SSIM = $0.77$, compression rate $\approx1393$). } 
\label{simulated}
\end{figure*}

We examine two training strategies: the first is our proposed \emph{data-to-image compression} which optimizes equation (\ref{loss}) using $\{ \mathbf f^{(i)}, \mathbf u^{(i)} \}_{i=1}^N$. The second strategy is a \emph{data-to-data compression} agnostic to DAS, which optimizes equation (\ref{loss}) but we replace its first term by $\sum_{i}\| \mathbf f^{(i)} - \mathcal{D}_{\boldsymbol \phi}(\mathcal E_{\boldsymbol \theta}(\mathbf f^{(i)}) ) \|_{2}^2$. It uses only $\{ \mathbf f^{(i)}\}_{i=1}^N$ as training data.

\section{Experiments}
To evaluate our proposed approach, we examine ultrasonic-based non-destructive inspection of pipelines. We simulated ultrasonic data using k-Wave \cite{kwave}. The number and location of defects in a pipeline were randomly varied, resulting in 770 scenarios (training: 700, test: 70). An example of a speed of sound map for a random scenario is shown in Figure \ref{simulated}(a). The pipeline was modelled as carbon steel with speed of sound, $s = 5920$ m/s, the defects and the pipe wall were modelled as air, $s = 343$ m/s. The data domain is $1020\times 64\times 64$ with 64 elements, 1020 time samples and 50 MHz sampling frequency. The image domain is set to $72\times 118$. The simulated data were used as input during training and the corresponding DAS images were used as targets. An example of a target image can be seen in Figure \ref{simulated}(b). Using this data simulation setup, the DAS operator is defined as $\mathcal B \in \mathbb{R}^{(72\times 118) \times (1020\times 64 \times 64)}$. Therefore, the best lossless linear compression would have compression rate of $\frac{1020\times 64\times 64}{72\times118} \approx 492$. 

\subsection*{Data-to-image vs data-to-data compression}
Using our proposed architecture, we can vary the compression rate by adding layers. We examine two compression rates, one that is close to the best lossless linear rate and one that is higher. We train our proposed data-to-image compression as well as a data-to-data compression, agnostic to subsequent image formation as discussed in the previous section.

The first experiment examines an encoder with 5 layers of 32 filters and a last layer (in total 6) with 256 filters. This results in a compressed code of $62\times12\times 12$ and gives a compression rate of $\frac{1020\times 64\times 64}{62\times 12\times 12} \approx 468$. At this compression rate, the average structural similarity (SSIM) index across the test set is: $0.93$ for our proposed data-to-image compression and $0.86$ for the data-to-data compression followed by DAS. Figure \ref{simulated}(c) and \ref{simulated}(d) show the decompressed images using our proposed data-to-image compression and the data-to-data compression. Both approaches produce high quality images.

The second experiment examines an encoder with 6 layers of 32 filters and a last layer (in total 7) with 256 filters. This results in a compressed code of $30\times 10\times 10$. This gives a compression rate of $\frac{1020\times 64\times 64}{30\times 10\times 10} \approx 1393$. At this compression rate, the average SSIM across the test set is: $0.91$ for our proposed data-to-image compression and $0.77$ for the data-to-data compression followed by DAS. Figure \ref{simulated}(e) shows the decompressed image using our proposed data-to-image compression tailored towards DAS. We can see that all defects are imaged with great accuracy. Figure \ref{simulated}(f) shows the decompressed image using data-to-data compression. One defect is correctly localized but the two others are not imaged accurately, deeming their precise localization challenging.

\section{Conclusion}
Deep learning is being utilized for more and more tasks in fast ultrasonic imaging, including for data compression in between acquisition and image formation. The compression rates obtained by generic compression schemes can be improved significantly if the goal of image formation is taken into account, and compression networks are trained end-to-end towards this objective. In this work, we proposed a novel encoder-decoder architecture that explicitly incorporates the DAS imaging operator as a network layer. This way, the network can exploit the approximations and reductions of acoustic wave-matter interactions performed by image formation. This achieves data-to-image compression, where the raw data is transformed into a compressed code through vector quantization that exploits patterns in acoustic waves via a codebook. Then, the data are decompressed directly into an image. We compared the proposed data-to-image compression against a data-to-data compression that optimizes weights without any information on the subsequent image formation method. Experiments show that our proposed data-to-image compression obtains much better image quality, for our application, at compression rates that are considerably higher than the theoretical best lossless linear compression rates. This illustrates the great potential of designing deep data compression methods tailored to an image formation method.

\bibliographystyle{IEEEtran}
\bibliography{references}

\end{document}